\documentclass{openjournal}

\usepackage{amsmath,amssymb}
\usepackage{natbib}
\usepackage{graphicx}
\usepackage{booktabs}
\usepackage{microtype}
\usepackage{xcolor}
\usepackage[utf8]{inputenc}
\usepackage[english]{babel}
\usepackage{hyperref}
\hypersetup{
    unicode,
    colorlinks=true,
    linkcolor=linkcolor,
    citecolor=linkcolor,
    filecolor=linkcolor,
    urlcolor=linkcolor
}
\usepackage{color,colortbl}
\definecolor{linkcolor}{rgb}{0.0,0.3,0.5}
\usepackage{tensind}
\tensordelimiter{?}
\DeclareGraphicsExtensions{.bmp,.png,.jpg,.pdf}
\usepackage{verbatim}
\usepackage[normalem]{ulem}
\usepackage{orcidlink}
\usepackage{soul}
\usepackage{longtable}

\usepackage{amsmath,amssymb}
\usepackage{natbib}
\usepackage{graphicx}
\usepackage{booktabs}
\usepackage{microtype}
\usepackage{xcolor}
\usepackage{hyperref}

\hypersetup{
  colorlinks=true,
  linkcolor=blue,
  citecolor=blue,
  urlcolor=blue
}

\begin{document}

\title{Tidally-Controlled Fragmentation around Black Holes, Massive Clumps, Protostars, and the Galactic Center}

\author{Guang-Xing Li}
\affiliation{South-Western Institute for Astronomy Research, Yunnan University, Kunming, 650500 Yunnan, P.R. China}
\email{ligx.ngc7293@gmail.com, gxli@ynu.edu.cn}
\begin{abstract}
Gravity plays important roles at multiple scales in the universe. An important,
yet often neglected, role of gravity is its ability in driving anisotropic
fragmentation through tides. When tides dominate, fragmentation becomes
anisotropic, and the Jeans length along the short axis, $l_{\rm tidal, Jeans}$,
is approximately $\sigma_{\rm v}/\sqrt{T} \approx \sigma_{\rm v}/\sqrt{G
\rho_{\rm mean}}$, determined by the external tides through the mean density
$\rho_{\rm mean}$. We compare predictions of $l_{\rm tidal, Jeans}$ against
observational results in massive star-forming clumps, the Circumnuclear Disk
(CND) around the supermassive black hole Sgr A* at the center of the Galaxy, the
Central Molecular Zone in the Galactic Center, a hub-filament system and a streamer
around a young star. We find that the observed widths of these filamentary
structures match theoretical predictions from tidally-controlled Jeans
fragmentation. The formation of filaments can potentially shield cold gas
against radiation pressure and photoevaporation, as well as hydrodynamical
interaction with the ambient medium, potentially enabling the cold gas to
survive. Thus, tidal forces are major players regulating gas transport around massive objects.
\end{abstract}

\section{Introduction}
Gravity controls the formation and evolution of a variety of systems in the
universe. Tides, which arise from the spatial difference in gravitational
acceleration, are an active process in numerous astrophysical systems. Around a
massive object, the gravitational acceleration is $a = G m r^{-2}$. Tides are
caused by the spatial difference in the gravitational acceleration, e.g.,
$T_{ij} = \frac{\partial a_i}{\partial x_j}$, where $x$ is a spatial coordinate.
The strength of the tides is of the order $T \approx G m r^{-3} \approx G
\rho_{\rm mean}$, where $\rho_{\rm mean}$ is the mean density of the system
measured from the nearest center to the location of interest. The dispersion
relation of gas under external tides has been derived by
\cite{2013MNRAS.434L..56J}. Tidal forces
dominate when $\rho_{\rm mean} > \rho_{\rm local}$, where $\rho_{\rm local}$ is
the local density of the system. The importance of tides lead 
\citep{2024MNRAS.528L..52L}, and its significance has been accepted ever since
\citep{2025A&A...696L...5Z}.

The Jeans length is the characteristic length scale of gravitational instability, representing the scale at which the sound crossing time ($t_{\rm cross} = l / \sigma_{\rm v}$) equals the free-fall time ($t_{\rm ff} = 1/\sqrt{G \rho_{\rm local}}$). Tidal effects dominate when the mean density of the central object observed at the location of interest is greater than the local density, i.e., $\rho_{\rm mean} > \rho_{\rm local}$. In this case, the Jeans length is modified by the tides \citep{2013MNRAS.434L..56J}. Since tides are anisotropic, the fragmentation is also anisotropic: it is longer along the radial direction and shorter along the azimuthal direction. This anisotropic fragmentation can lead to the formation of filamentary structures around massive objects \citep{2024MNRAS.532.1126L}. These filaments are often characterized by their widths, which, under tidally-driven Jeans fragmentation, is given by:
\begin{equation}
l_{\rm tidal, Jeans} = \frac{\sigma_{\rm v}}{\sqrt{T}} = \frac{\sigma_{\rm v}}{\sqrt{G \rho_{\rm mean}}},
\label{eq:t:jeans}
\end{equation}
where $\rho_{\rm mean}$ is the mean density of the central object measured at distance $r$.

Accretion is a common process leading to the growth of various objects in the universe, and filamentary structures are often observed in the accretion flow around dense objects. Examples of this include streamers around young stars \citep{2020NatAs...4.1158P}, filamentary structures in massive star-forming regions \citep{2018A&A...610A..77H}, and the circumnuclear disk around Sgr A* \citep{2021ApJ...913...94H}. In this paper, we will show that the observed widths of these filamentary structures match theoretical predictions from tidal fragmentation, indicating a tidal origin for these filamentary structures.

\section{Methods and Results}
We collected literature values for the widths of filamentary structures, and the results are listed in Table \ref{tab:width}. The sample includes filaments in star-forming regions, such as OMC-1 and DR21 \citep{2012A&A...543L...3H,2018A&A...610A..77H}, the Central Molecular Zone (CMZ), a gaseous structure of approximately 200 pc \citep{Bally1987}, and the Circumnuclear Disk (CND), a gaseous structure of approximately 3 pc \citep{1998MNRAS.294...35S}, as well as a gas streamer around a young protostar \citep{2020NatAs...4.1158P}. These are either values quoted by the authors, as in the case of the fibers in OMC-1 \citep{2018A&A...610A..77H}, or values estimated using, e.g., the size of the clumps, as in the case of the CMZ \citep{2020ApJ...897...89L}, or values measured from published figures when the data are not easily accessible, as in the case of the streamer around IRAS 03292+3039 \citep{2020NatAs...4.1158P} and the CND \citep{2021ApJ...913...94H}. These are well-known systems where filamentary structures are observed around massive objects, such that gravity from the central object is expected to dominate. The details on how these values are derived can be found in Appendix \ref{sec:appendix}.

The sizes of the systems range from 5000 AU (streamer) to 200 pc (CMZ), and their density span approximately five orders of magnitude. The source of gravity also varies: in cases related to star formation, e.g., the OMC-1 and DR21 star-forming regions \citep{2012A&A...543L...3H,2018A&A...610A..77H}, as well as the gas streamer around the protostar \citep{2020NatAs...4.1158P}, the source of gravity is the central object. In the CMZ \citep{Bally1987,Morris1996}, gravity originates from stars in the Galactic Bulge, and in the CND \citep{1998MNRAS.294...35S}, gravity originates from the Sgr A* supermassive black hole. The diversity of these systems implies that understanding the importance of tides in these regions can improve our understanding of multiple areas in modern astrophysics, including star formation, galaxy centers, and the growth of supermassive black holes.

\begin{figure*}[h!]
\centering
\includegraphics[width=\textwidth]{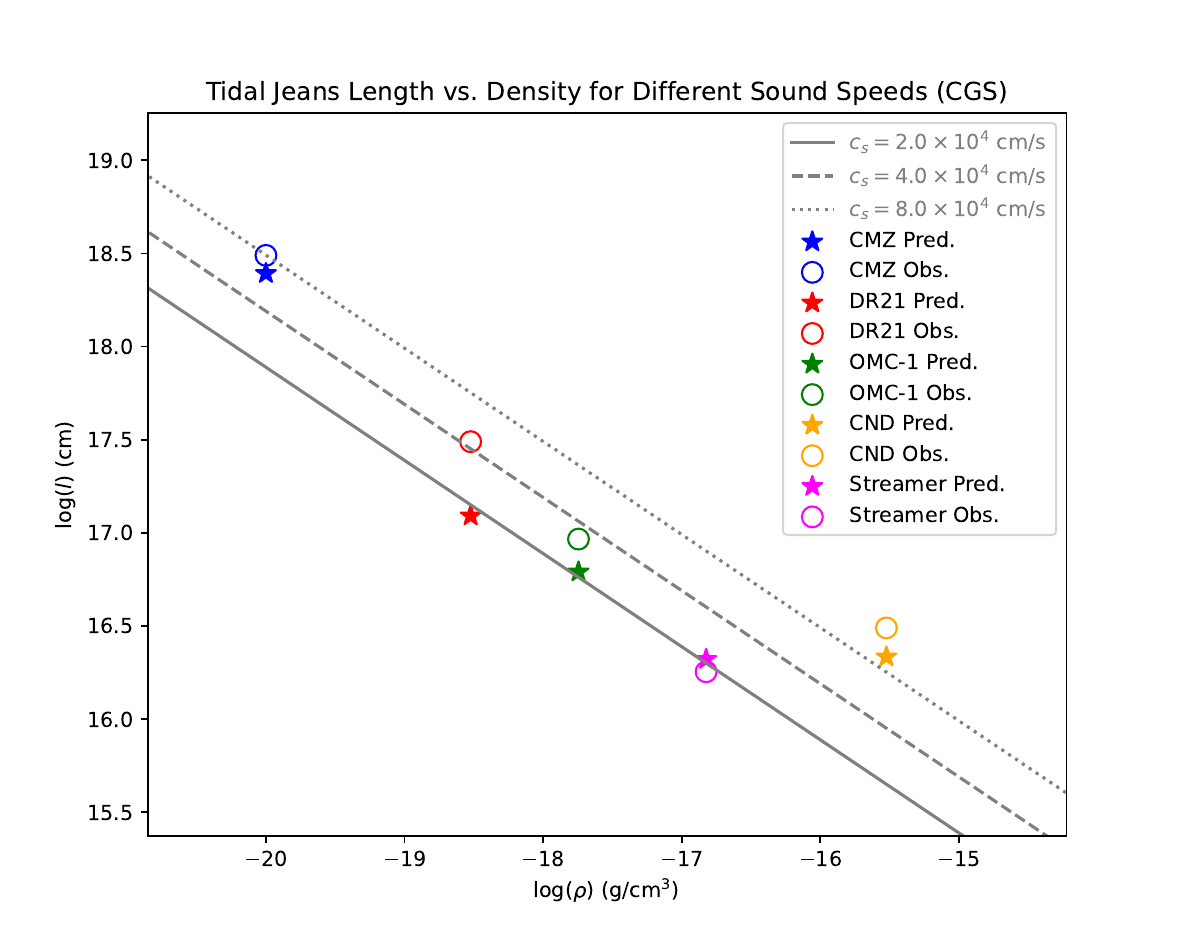}
\caption{Comparison between the tidal Jeans length (theoretical prediction) and the observed width of filamentary structures. The data points correspond to the values listed in Table \ref{tab:width}. \label{fig:width_comparison}}
\end{figure*}

In Figure \ref{fig:width_comparison}, we present a comparison between the tidal
Jeans length and the observed widths of filamentary structures, where the
results appear to agree. We also observe some streams that are narrower than the
predicted value, indicating that these streams might have multiple origins. In
the circumnuclear disk around Sgr A*, the observed width is larger than the
predicted value (e.g., DR21, Orion A, and CND). There are a few reasons for this
discrepancy. First, we are performing order-of-magnitude comparisons, where we
observe a typical difference of a factor of few, as expected from the
semi-quantitative nature of the analysis. What is remarkable, however, is the
closeness of the measured and predicted values, given the diversity of the
systems measured in change of the density which spans 5 orders of magnitudes.
We note that in
the case of the CND, the observed structures tend to pile up at small radii
\citep{2021ApJ...913...94H}, hinting at insufficient resolution, where the
fragmentation length from observation is likely an upper limit.

\begin{deluxetable*}{ccccccc}
\tablecaption{Comparison of Filamentary Structure Widths with Theoretical Predictions \label{tab:width}}
\tablehead{
\colhead{Name} & \colhead{Size} & \colhead{$\rho_{\rm mean}$ (g cm$^{-3}$)} & \colhead{$\sigma_{\rm v}$ (km s$^{-1}$)} & \colhead{Tidal Jeans Length} & \colhead{Filament Width} & \colhead{Source of Tides}
}
\startdata
Streamer & 5000 AU & $1.5 \times 10^{-17}$ & 0.2 & 1400 AU & 1200 AU & Central Star \\
OMC-1 & 0.5 pc & $1.8 \times 10^{-18}$ & 0.2 & 0.02 pc & 0.03 pc & Gas at Center \\
DR21 & 1.5 pc & $3 \times 10^{-19}$ & 0.2 & 0.04 pc & 0.1 pc & Gas at Center \\
CND & 1 pc & $3 \times 10^{-16}$ & 1 & 0.007 pc & 0.01 pc & SgrA* Black Hole \\
CMZ & 120 pc & $10^{-20}$ & 0.6 & 0.8 pc & 1 pc & Stellar Bulge \\
\enddata
\tablecomments{Values are taken from the literature (see Appendix).}
\end{deluxetable*}

\section{Conclusion}
We offer a tidal fragmentation explanation for the filamentary structures observed in a variety of astrophysical systems, including the CMZ, the CND around Sgr A*, the fibers in OMC-1, and streamers around young stars. The sizes of these systems range from 5000 AU (gas streamer) to 120 pc (CND), and their density vary over four orders of magnitude. In all these systems, we observe the existence of thin, filamentary structures.

Our analysis shows that in these systems, the existence of filamentary structures results from tidal Jeans fragmentation. Since tides are a natural consequence of gravity and are activated in many astrophysical systems, they offer a unifying explanation for these filamentary structures.

The fact that tides can dominate has implications for the evolution of the gas
and the growth of the central object. When tides dominate, fragmentation is
anisotropic, leading to the formation of filamentary structures. This spatial
organization has important implications for the accretion process. The first is
the ubiquitous existence of filaments around many systems. For examples,
hub-filament systems are common in massive star-forming regions
\citep{}. 

The first is the reduction of exposure to the radiation field from nearby objects. A frequently discussed barrier to accretion is radiation pressure, as well as the photoevaporation effect from the radiation field. This issue is discussed both in the star formation community as the radiation barrier \citep{2005IAUS..227..231K,2015ApJ...815...68M}, and in the black hole community under the term "Eddington limit" \citep{Eddington1926}, where arranging the gas into columns is an effective way to reduce exposure to the radiation field. Tidally-regulated fragmentation is one possible way to form these well-organized filament systems.

Another barrier to mass transport is the hydrodynamical interaction \citep{2019AJ....158..124F} between the cold accretion flow and the ambient medium. Thermal conduction and Kelvin-Helmholtz instability can lead to mixing between the cold and hot gas, which might dissolve the cold gas. Studies already show that arranging gas into filaments is an effective way to reduce the interaction between the cold and hot gas, ensuring the survival of the cold gas, where tidally-regulated fragmentation is also a possible way to achieve this filamentary configuration. Apart from driving the formation of filaments, tidally-induced Jeans fragmentation might play an important role in maintaining gas accretion onto, e.g., dense molecular clumps, the CND around Sgr A*, and in the CMZ.

\appendix
\section{Literature Values}\label{sec:appendix}
Values quoted in Table \ref{tab:width} are derived from observations presented in the literature. The derivations are listed below.

\subsection{Fibers in OMC-1}
We adopt a total mass of 3000 $M_\odot$ \citep{2019A&A...622A..91G}, and at a fiducial distance of 0.5 pc where the "fibers" are observed, the density is $\rho \approx m r^{-3} \approx 1.8 \times 10^{-18}$ g cm$^{-3}$. Adopting a temperature of 10 K, we derive a thermal sound speed of 0.2 km s$^{-1}$ for the dense molecular gas, where
\begin{equation}
l_{\rm Jeans, Tidal} = \frac{\sigma_{\rm v}}{\sqrt{G \rho}} = 0.02 \, \text{pc},
\end{equation}
which is consistent with the observed width of 0.03 pc \citep{2018A&A...610A..77H}.

\subsection{Fibers around the DR21 Filament}
The DR21 filament is a well-studied star-forming region with a set of filamentary structures extending from the central spine, which is composed of a number of clumps. The total mass of the DR21 central region is estimated to be 15000 $M_\odot$ \citep{2012A&A...543L...3H}, and adopting a distance of 1.5 pc measured from the map, we find that the density is $3 \times 10^{-19}$ g cm$^{-3}$. Assuming a thermal sound speed of 0.2 km s$^{-1}$, we have
\begin{equation}
l_{\rm Jeans, Tidal} = \frac{\sigma_{\rm v}}{\sqrt{G \rho}} = 0.04 \, \text{pc},
\end{equation}
which is comparable to the width of 0.1 pc \citep{2022ApJ...927..106C} measured using high-resolution interferometer observations.

\subsection{CMZ}
The Central Molecular Zone (CMZ) \citep{Bally1987} is a disk-like gas structure that rotates around the center of the Galaxy. The region has a radius of 250 pc and contains a total of $3 \times 10^7 M_{\odot}$ of molecular gas. The effect of shear in shaping the gas density of the CMZ has been discussed \citep{2020ApJ...897...89L}, where we estimate, assuming a circular rotation,
\begin{equation}
\rho_{\rm mean} = \frac{v^2}{G r^2} = \frac{(100 \, \text{km s}^{-1})^2}{G (125 \, \text{pc})^2} = 10^{-20} \, \text{g cm}^{-3},
\end{equation}
where we have assumed a rotational velocity of 100 km s$^{-1}$ \citep{2022MNRAS.516..907S}.

Gas in the CMZ is warmer than the local interstellar medium, with a temperature of around 80 K \citep{2016A&A...586A..50G}. This leads to a sound speed of $\sigma_{\rm v} \approx \sqrt{k T / m_{H_2}} = 0.6 \, \text{km s}^{-1}$, from which we have
\begin{equation}
l = \frac{\sigma_{\rm v}}{\sqrt{G \rho}} = 0.8 \, \text{pc}.
\end{equation}

Gas in the CMZ is organized in streamers \citep{2016MNRAS.457.2675H}. We use the typical size of the gas clumps \citep{2020ApJ...897...89L} to estimate the width of the streamers, from which we find
\begin{equation}
l_{\rm streamers} \approx 1 \, \text{pc},
\end{equation}
consistent with estimates using the tidal-Jeans formalism.

\subsection{CND around SgrA*}
The circumnuclear disk (CND) around SgrA* is a disk-like structure that rotates around the central black hole of the Milky Way. It contains a mass between $10^{4-6} M_\odot$ and a size of around 2 pc \citep{2005ApJ...622..346C,2012A&A...542L..21R}.

To compute the strength of the tidal force, we note that the mass of the central black hole is around $4 \times 10^6 M_{\odot}$, and at 1 pc, the mean density is
\begin{equation}
\rho_{\rm mean} = \frac{M_{\rm BH}}{r^3} = \frac{4 \times 10^6 M_\odot}{(1 \, \text{pc})^3} \approx 3 \times 10^{-16} \, \text{g cm}^{-3} \approx 10^8 m_{\rm H_2}.
\end{equation}
Observations reveal two populations of clumps, with the hotter populations having smaller density and being tidally dominated \citep{2021ApJ...913...94H}.

To compute the fragmentation length, assuming a temperature of 300 K \citep{2021ApJ...913...94H}, we have $c_{\rm s} \approx 1 \, \text{km s}^{-1}$, thus
\begin{equation}
l = \frac{\sigma_{\rm v}}{\sqrt{G \rho}} = 0.007 \, \text{pc},
\end{equation}
which is comparable, but still smaller than the reported minimum clump size of $\gtrsim 0.01$ pc. The fact that the ALMA observations have revealed a large number of filamentary structures is consistent with this tidal interpretation. Since there is a clear pile-up of cores towards the smaller scales due to the limited resolution of the observations, the size difference is not a real discrepancy but a reflection of observational limits.

\subsection{Streamer around IRAS 03292+3039}
Recent observations have reported the existence of streamers around young stars. These are filamentary structures that are accreting onto the star and are often found in isolation. Their elongated morphology makes the tidal effect a strong candidate to explain their formation. In IRAS 03292+3039 \citep{2020NatAs...4.1158P}, the authors found a stream around a young protostar of $3 M_\odot$, at a distance of 5000 AU. The mean density is
\begin{equation}
\rho_{\rm mean} = 1.5 \times 10^{-17} \, \text{g cm}^{-3},
\end{equation}
and assuming a sound speed of 0.2 km s$^{-1}$, we have
\begin{equation}
l = \frac{\sigma_{\rm v}}{\sqrt{G \rho_{\rm mean}}} \approx 1400 \, \text{AU},
\end{equation}
and from the published figures, we estimate a width of 4 arcsec, which corresponds to 1200 AU, consistent with the analytical estimations.

\section*{Acknowledgments}
GXL acknowledges support from NSFC grant No. 12273032 and 12033005.
\bibliographystyle{apsrev4-1}
\bibliography{paper}

\begin{thebibliography}{23}%
\makeatletter
\providecommand \@ifxundefined [1]{%
 \@ifx{#1\undefined}
}%
\providecommand \@ifnum [1]{%
 \ifnum #1\expandafter \@firstoftwo
 \else \expandafter \@secondoftwo
 \fi
}%
\providecommand \@ifx [1]{%
 \ifx #1\expandafter \@firstoftwo
 \else \expandafter \@secondoftwo
 \fi
}%
\providecommand \natexlab [1]{#1}%
\providecommand \enquote  [1]{``#1''}%
\providecommand \bibnamefont  [1]{#1}%
\providecommand \bibfnamefont [1]{#1}%
\providecommand \citenamefont [1]{#1}%
\providecommand \href@noop [0]{\@secondoftwo}%
\providecommand \href [0]{\begingroup \@sanitize@url \@href}%
\providecommand \@href[1]{\@@startlink{#1}\@@href}%
\providecommand \@@href[1]{\endgroup#1\@@endlink}%
\providecommand \@sanitize@url [0]{\catcode `\\12\catcode `\$12\catcode `\&12\catcode `\#12\catcode `\^12\catcode `\_12\catcode `\%12\relax}%
\providecommand \@@startlink[1]{}%
\providecommand \@@endlink[0]{}%
\providecommand \url  [0]{\begingroup\@sanitize@url \@url }%
\providecommand \@url [1]{\endgroup\@href {#1}{\urlprefix }}%
\providecommand \urlprefix  [0]{URL }%
\providecommand \Eprint [0]{\href }%
\providecommand \doibase [0]{http://dx.doi.org/}%
\providecommand \selectlanguage [0]{\@gobble}%
\providecommand \bibinfo  [0]{\@secondoftwo}%
\providecommand \bibfield  [0]{\@secondoftwo}%
\providecommand \translation [1]{[#1]}%
\providecommand \BibitemOpen [0]{}%
\providecommand \bibitemStop [0]{}%
\providecommand \bibitemNoStop [0]{.\EOS\space}%
\providecommand \EOS [0]{\spacefactor3000\relax}%
\providecommand \BibitemShut  [1]{\csname bibitem#1\endcsname}%
\let\auto@bib@innerbib\@empty
\bibitem [{\citenamefont {{Jog}}(2013)}]{2013MNRAS.434L..56J}%
  \BibitemOpen
  \bibfield  {author} {\bibinfo {author} {\bibfnamefont {C.~J.}\ \bibnamefont {{Jog}}},\ }\href {\doibase 10.1093/mnrasl/slt077} {\bibfield  {journal} {\bibinfo  {journal} {\mnras}\ }\textbf {\bibinfo {volume} {434}},\ \bibinfo {pages} {L56} (\bibinfo {year} {2013})},\ \Eprint {http://arxiv.org/abs/1306.4425} {arXiv:1306.4425 [astro-ph.GA]} \BibitemShut {NoStop}%
\bibitem [{\citenamefont {{Li}}(2024{\natexlab{a}})}]{2024MNRAS.528L..52L}%
  \BibitemOpen
  \bibfield  {author} {\bibinfo {author} {\bibfnamefont {G.-X.}\ \bibnamefont {{Li}}},\ }\href {\doibase 10.1093/mnrasl/slad149} {\bibfield  {journal} {\bibinfo  {journal} {\mnras}\ }\textbf {\bibinfo {volume} {528}},\ \bibinfo {pages} {L52} (\bibinfo {year} {2024}{\natexlab{a}})},\ \Eprint {http://arxiv.org/abs/2309.16125} {arXiv:2309.16125 [astro-ph.GA]} \BibitemShut {NoStop}%
\bibitem [{\citenamefont {{Zhou}}(2025)}]{2025A&A...696L...5Z}%
  \BibitemOpen
  \bibfield  {author} {\bibinfo {author} {\bibfnamefont {J.~W.}\ \bibnamefont {{Zhou}}},\ }\href {\doibase 10.1051/0004-6361/202453040} {\bibfield  {journal} {\bibinfo  {journal} {\aap}\ }\textbf {\bibinfo {volume} {696}},\ \bibinfo {eid} {L5} (\bibinfo {year} {2025})},\ \Eprint {http://arxiv.org/abs/2503.16937} {arXiv:2503.16937 [astro-ph.GA]} \BibitemShut {NoStop}%
\bibitem [{\citenamefont {{Li}}(2024{\natexlab{b}})}]{2024MNRAS.532.1126L}%
  \BibitemOpen
  \bibfield  {author} {\bibinfo {author} {\bibfnamefont {G.-X.}\ \bibnamefont {{Li}}},\ }\href {\doibase 10.1093/mnras/stae900} {\bibfield  {journal} {\bibinfo  {journal} {\mnras}\ }\textbf {\bibinfo {volume} {532}},\ \bibinfo {pages} {1126} (\bibinfo {year} {2024}{\natexlab{b}})},\ \Eprint {http://arxiv.org/abs/2403.02612} {arXiv:2403.02612 [astro-ph.GA]} \BibitemShut {NoStop}%
\bibitem [{\citenamefont {{Pineda}}\ \emph {et~al.}(2020)\citenamefont {{Pineda}}, \citenamefont {{Segura-Cox}}, \citenamefont {{Caselli}}, \citenamefont {{Cunningham}}, \citenamefont {{Zhao}}, \citenamefont {{Schmiedeke}}, \citenamefont {{Maureira}},\ and\ \citenamefont {{Neri}}}]{2020NatAs...4.1158P}%
  \BibitemOpen
  \bibfield  {author} {\bibinfo {author} {\bibfnamefont {J.~E.}\ \bibnamefont {{Pineda}}}, \bibinfo {author} {\bibfnamefont {D.}~\bibnamefont {{Segura-Cox}}}, \bibinfo {author} {\bibfnamefont {P.}~\bibnamefont {{Caselli}}}, \bibinfo {author} {\bibfnamefont {N.}~\bibnamefont {{Cunningham}}}, \bibinfo {author} {\bibfnamefont {B.}~\bibnamefont {{Zhao}}}, \bibinfo {author} {\bibfnamefont {A.}~\bibnamefont {{Schmiedeke}}}, \bibinfo {author} {\bibfnamefont {M.~J.}\ \bibnamefont {{Maureira}}}, \ and\ \bibinfo {author} {\bibfnamefont {R.}~\bibnamefont {{Neri}}},\ }\href {\doibase 10.1038/s41550-020-1150-z} {\bibfield  {journal} {\bibinfo  {journal} {Nature Astronomy}\ }\textbf {\bibinfo {volume} {4}},\ \bibinfo {pages} {1158} (\bibinfo {year} {2020})},\ \Eprint {http://arxiv.org/abs/2007.13430} {arXiv:2007.13430 [astro-ph.GA]} \BibitemShut {NoStop}%
\bibitem [{\citenamefont {{Hacar}}\ \emph {et~al.}(2018)\citenamefont {{Hacar}}, \citenamefont {{Tafalla}}, \citenamefont {{Forbrich}}, \citenamefont {{Alves}}, \citenamefont {{Meingast}}, \citenamefont {{Grossschedl}},\ and\ \citenamefont {{Teixeira}}}]{2018A&A...610A..77H}%
  \BibitemOpen
  \bibfield  {author} {\bibinfo {author} {\bibfnamefont {A.}~\bibnamefont {{Hacar}}}, \bibinfo {author} {\bibfnamefont {M.}~\bibnamefont {{Tafalla}}}, \bibinfo {author} {\bibfnamefont {J.}~\bibnamefont {{Forbrich}}}, \bibinfo {author} {\bibfnamefont {J.}~\bibnamefont {{Alves}}}, \bibinfo {author} {\bibfnamefont {S.}~\bibnamefont {{Meingast}}}, \bibinfo {author} {\bibfnamefont {J.}~\bibnamefont {{Grossschedl}}}, \ and\ \bibinfo {author} {\bibfnamefont {P.~S.}\ \bibnamefont {{Teixeira}}},\ }\href {\doibase 10.1051/0004-6361/201731894} {\bibfield  {journal} {\bibinfo  {journal} {\aap}\ }\textbf {\bibinfo {volume} {610}},\ \bibinfo {eid} {A77} (\bibinfo {year} {2018})},\ \Eprint {http://arxiv.org/abs/1801.01500} {arXiv:1801.01500 [astro-ph.GA]} \BibitemShut {NoStop}%
\bibitem [{\citenamefont {{Hsieh}}\ \emph {et~al.}(2021)\citenamefont {{Hsieh}}, \citenamefont {{Koch}}, \citenamefont {{Kim}}, \citenamefont {{Mart{\'\i}n}}, \citenamefont {{Yen}}, \citenamefont {{Carpenter}}, \citenamefont {{Harada}}, \citenamefont {{Turner}}, \citenamefont {{Ho}}, \citenamefont {{Tang}},\ and\ \citenamefont {{Beck}}}]{2021ApJ...913...94H}%
  \BibitemOpen
  \bibfield  {author} {\bibinfo {author} {\bibfnamefont {P.-Y.}\ \bibnamefont {{Hsieh}}}, \bibinfo {author} {\bibfnamefont {P.~M.}\ \bibnamefont {{Koch}}}, \bibinfo {author} {\bibfnamefont {W.-T.}\ \bibnamefont {{Kim}}}, \bibinfo {author} {\bibfnamefont {S.}~\bibnamefont {{Mart{\'\i}n}}}, \bibinfo {author} {\bibfnamefont {H.-W.}\ \bibnamefont {{Yen}}}, \bibinfo {author} {\bibfnamefont {J.~M.}\ \bibnamefont {{Carpenter}}}, \bibinfo {author} {\bibfnamefont {N.}~\bibnamefont {{Harada}}}, \bibinfo {author} {\bibfnamefont {J.~L.}\ \bibnamefont {{Turner}}}, \bibinfo {author} {\bibfnamefont {P.~T.~P.}\ \bibnamefont {{Ho}}}, \bibinfo {author} {\bibfnamefont {Y.-W.}\ \bibnamefont {{Tang}}}, \ and\ \bibinfo {author} {\bibfnamefont {S.}~\bibnamefont {{Beck}}},\ }\href {\doibase 10.3847/1538-4357/abf4cd} {\bibfield  {journal} {\bibinfo  {journal} {\apj}\ }\textbf {\bibinfo {volume} {913}},\ \bibinfo {eid} {94} (\bibinfo {year} {2021})},\ \Eprint {http://arxiv.org/abs/2104.02078} {arXiv:2104.02078 [astro-ph.GA]} \BibitemShut {NoStop}%
\bibitem [{\citenamefont {{Hennemann}}\ \emph {et~al.}(2012)\citenamefont {{Hennemann}}, \citenamefont {{Motte}}, \citenamefont {{Schneider}}, \citenamefont {{Didelon}}, \citenamefont {{Hill}}, \citenamefont {{Arzoumanian}}, \citenamefont {{Bontemps}}, \citenamefont {{Csengeri}}, \citenamefont {{Andr{\'e}}}, \citenamefont {{Konyves}}, \citenamefont {{Louvet}}, \citenamefont {{Marston}}, \citenamefont {{Men'shchikov}}, \citenamefont {{Minier}}, \citenamefont {{Nguyen Luong}}, \citenamefont {{Palmeirim}}, \citenamefont {{Peretto}}, \citenamefont {{Sauvage}}, \citenamefont {{Zavagno}}, \citenamefont {{Anderson}}, \citenamefont {{Bernard}}, \citenamefont {{Di Francesco}}, \citenamefont {{Elia}}, \citenamefont {{Li}}, \citenamefont {{Martin}}, \citenamefont {{Molinari}}, \citenamefont {{Pezzuto}}, \citenamefont {{Russeil}}, \citenamefont {{Rygl}}, \citenamefont {{Schisano}}, \citenamefont {{Spinoglio}}, \citenamefont {{Sousbie}}, \citenamefont {{Ward-Thompson}},\ and\ \citenamefont {{White}}}]{2012A&A...543L...3H}%
  \BibitemOpen
  \bibfield  {author} {\bibinfo {author} {\bibfnamefont {M.}~\bibnamefont {{Hennemann}}}, \bibinfo {author} {\bibfnamefont {F.}~\bibnamefont {{Motte}}}, \bibinfo {author} {\bibfnamefont {N.}~\bibnamefont {{Schneider}}}, \bibinfo {author} {\bibfnamefont {P.}~\bibnamefont {{Didelon}}}, \bibinfo {author} {\bibfnamefont {T.}~\bibnamefont {{Hill}}}, \bibinfo {author} {\bibfnamefont {D.}~\bibnamefont {{Arzoumanian}}}, \bibinfo {author} {\bibfnamefont {S.}~\bibnamefont {{Bontemps}}}, \bibinfo {author} {\bibfnamefont {T.}~\bibnamefont {{Csengeri}}}, \bibinfo {author} {\bibfnamefont {P.}~\bibnamefont {{Andr{\'e}}}}, \bibinfo {author} {\bibfnamefont {V.}~\bibnamefont {{Konyves}}}, \bibinfo {author} {\bibfnamefont {F.}~\bibnamefont {{Louvet}}}, \bibinfo {author} {\bibfnamefont {A.}~\bibnamefont {{Marston}}}, \bibinfo {author} {\bibfnamefont {A.}~\bibnamefont {{Men'shchikov}}}, \bibinfo {author} {\bibfnamefont {V.}~\bibnamefont {{Minier}}}, \bibinfo {author} {\bibfnamefont {Q.}~\bibnamefont {{Nguyen Luong}}}, \bibinfo {author} {\bibfnamefont {P.}~\bibnamefont {{Palmeirim}}}, \bibinfo {author} {\bibfnamefont {N.}~\bibnamefont {{Peretto}}}, \bibinfo {author} {\bibfnamefont {M.}~\bibnamefont {{Sauvage}}}, \bibinfo {author} {\bibfnamefont {A.}~\bibnamefont {{Zavagno}}}, \bibinfo {author} {\bibfnamefont {L.~D.}\ \bibnamefont {{Anderson}}}, \bibinfo {author} {\bibfnamefont {J.~P.}\ \bibnamefont {{Bernard}}}, \bibinfo {author} {\bibfnamefont {J.}~\bibnamefont {{Di Francesco}}}, \bibinfo {author} {\bibfnamefont {D.}~\bibnamefont {{Elia}}}, \bibinfo {author} {\bibfnamefont {J.~Z.}\ \bibnamefont {{Li}}}, \bibinfo {author} {\bibfnamefont {P.~G.}\ \bibnamefont {{Martin}}}, \bibinfo {author} {\bibfnamefont {S.}~\bibnamefont {{Molinari}}}, \bibinfo {author} {\bibfnamefont {S.}~\bibnamefont {{Pezzuto}}}, \bibinfo {author} {\bibfnamefont {D.}~\bibnamefont {{Russeil}}}, \bibinfo {author} {\bibfnamefont {K.~L.~J.}\ \bibnamefont {{Rygl}}}, \bibinfo {author} {\bibfnamefont {E.}~\bibnamefont {{Schisano}}}, \bibinfo {author} {\bibfnamefont {L.}~\bibnamefont {{Spinoglio}}}, \bibinfo {author} {\bibfnamefont {T.}~\bibnamefont {{Sousbie}}}, \bibinfo {author} {\bibfnamefont {D.}~\bibnamefont {{Ward-Thompson}}}, \ and\ \bibinfo {author} {\bibfnamefont {G.~J.}\ \bibnamefont {{White}}},\ }\href {\doibase 10.1051/0004-6361/201219429} {\bibfield  {journal} {\bibinfo  {journal} {\aap}\ }\textbf {\bibinfo {volume} {543}},\ \bibinfo {eid} {L3} (\bibinfo {year} {2012})},\ \Eprint {http://arxiv.org/abs/1206.1243} {arXiv:1206.1243 [astro-ph.GA]} \BibitemShut {NoStop}%
\bibitem [{\citenamefont {{Bally}}\ \emph {et~al.}(1987)\citenamefont {{Bally}}, \citenamefont {{Stark}}, \citenamefont {{Wilson}},\ and\ \citenamefont {{Henkel}}}]{Bally1987}%
  \BibitemOpen
  \bibfield  {author} {\bibinfo {author} {\bibfnamefont {J.}~\bibnamefont {{Bally}}}, \bibinfo {author} {\bibfnamefont {A.~A.}\ \bibnamefont {{Stark}}}, \bibinfo {author} {\bibfnamefont {R.~W.}\ \bibnamefont {{Wilson}}}, \ and\ \bibinfo {author} {\bibfnamefont {C.}~\bibnamefont {{Henkel}}},\ }\href {\doibase 10.1086/191216} {\bibfield  {journal} {\bibinfo  {journal} {Astrophysical Journal Supplement Series}\ }\textbf {\bibinfo {volume} {65}},\ \bibinfo {pages} {13} (\bibinfo {year} {1987})}\BibitemShut {NoStop}%
\bibitem [{\citenamefont {{Sanders}}(1998)}]{1998MNRAS.294...35S}%
  \BibitemOpen
  \bibfield  {author} {\bibinfo {author} {\bibfnamefont {R.~H.}\ \bibnamefont {{Sanders}}},\ }\href {\doibase 10.1046/j.1365-8711.1998.01127.x} {\bibfield  {journal} {\bibinfo  {journal} {\mnras}\ }\textbf {\bibinfo {volume} {294}},\ \bibinfo {pages} {35} (\bibinfo {year} {1998})},\ \Eprint {http://arxiv.org/abs/astro-ph/9708272} {arXiv:astro-ph/9708272 [astro-ph]} \BibitemShut {NoStop}%
\bibitem [{\citenamefont {{Li}}\ and\ \citenamefont {{Zhang}}(2020)}]{2020ApJ...897...89L}%
  \BibitemOpen
  \bibfield  {author} {\bibinfo {author} {\bibfnamefont {G.-X.}\ \bibnamefont {{Li}}}\ and\ \bibinfo {author} {\bibfnamefont {C.-P.}\ \bibnamefont {{Zhang}}},\ }\href {\doibase 10.3847/1538-4357/ab8c47} {\bibfield  {journal} {\bibinfo  {journal} {\apj}\ }\textbf {\bibinfo {volume} {897}},\ \bibinfo {eid} {89} (\bibinfo {year} {2020})},\ \Eprint {http://arxiv.org/abs/1910.05015} {arXiv:1910.05015 [astro-ph.GA]} \BibitemShut {NoStop}%
\bibitem [{\citenamefont {Morris}\ and\ \citenamefont {Serabyn}(1996)}]{Morris1996}%
  \BibitemOpen
  \bibfield  {author} {\bibinfo {author} {\bibfnamefont {M.}~\bibnamefont {Morris}}\ and\ \bibinfo {author} {\bibfnamefont {E.}~\bibnamefont {Serabyn}},\ }\href {\doibase 10.1146/annurev.astro.34.1.645} {\bibfield  {journal} {\bibinfo  {journal} {Annual Review of Astronomy and Astrophysics}\ }\textbf {\bibinfo {volume} {34}},\ \bibinfo {pages} {645} (\bibinfo {year} {1996})},\ \bibinfo {note} {accessed: 2025-04-30}\BibitemShut {NoStop}%
\bibitem [{\citenamefont {{Krumholz}}\ \emph {et~al.}(2005)\citenamefont {{Krumholz}}, \citenamefont {{Klein}},\ and\ \citenamefont {{McKee}}}]{2005IAUS..227..231K}%
  \BibitemOpen
  \bibfield  {author} {\bibinfo {author} {\bibfnamefont {M.~R.}\ \bibnamefont {{Krumholz}}}, \bibinfo {author} {\bibfnamefont {R.~I.}\ \bibnamefont {{Klein}}}, \ and\ \bibinfo {author} {\bibfnamefont {C.~F.}\ \bibnamefont {{McKee}}},\ }in\ \href {\doibase 10.1017/S1743921305004588} {\emph {\bibinfo {booktitle} {Massive Star Birth: A Crossroads of Astrophysics}}},\ \bibinfo {series} {IAU Symposium}, Vol.\ \bibinfo {volume} {227},\ \bibinfo {editor} {edited by\ \bibinfo {editor} {\bibfnamefont {R.}~\bibnamefont {{Cesaroni}}}, \bibinfo {editor} {\bibfnamefont {M.}~\bibnamefont {{Felli}}}, \bibinfo {editor} {\bibfnamefont {E.}~\bibnamefont {{Churchwell}}}, \ and\ \bibinfo {editor} {\bibfnamefont {M.}~\bibnamefont {{Walmsley}}}}\ (\bibinfo {year} {2005})\ pp.\ \bibinfo {pages} {231--236},\ \Eprint {http://arxiv.org/abs/astro-ph/0510432} {arXiv:astro-ph/0510432 [astro-ph]} \BibitemShut {NoStop}%
\bibitem [{\citenamefont {{Matzner}}\ and\ \citenamefont {{Jumper}}(2015)}]{2015ApJ...815...68M}%
  \BibitemOpen
  \bibfield  {author} {\bibinfo {author} {\bibfnamefont {C.~D.}\ \bibnamefont {{Matzner}}}\ and\ \bibinfo {author} {\bibfnamefont {P.~H.}\ \bibnamefont {{Jumper}}},\ }\href {\doibase 10.1088/0004-637X/815/1/68} {\bibfield  {journal} {\bibinfo  {journal} {\apj}\ }\textbf {\bibinfo {volume} {815}},\ \bibinfo {eid} {68} (\bibinfo {year} {2015})},\ \Eprint {http://arxiv.org/abs/1511.03269} {arXiv:1511.03269 [astro-ph.GA]} \BibitemShut {NoStop}%
\bibitem [{\citenamefont {Eddington}(1926)}]{Eddington1926}%
  \BibitemOpen
  \bibfield  {author} {\bibinfo {author} {\bibfnamefont {A.~S.}\ \bibnamefont {Eddington}},\ }\href {\doibase 10.1017/CBO9780511608216} {\emph {\bibinfo {title} {The Internal Constitution of the Stars}}}\ (\bibinfo  {publisher} {Cambridge University Press},\ \bibinfo {address} {Cambridge, UK},\ \bibinfo {year} {1926})\ \bibinfo {note} {accessed: 2025-04-30}\BibitemShut {NoStop}%
\bibitem [{\citenamefont {{Forbes}}\ and\ \citenamefont {{Lin}}(2019)}]{2019AJ....158..124F}%
  \BibitemOpen
  \bibfield  {author} {\bibinfo {author} {\bibfnamefont {J.~C.}\ \bibnamefont {{Forbes}}}\ and\ \bibinfo {author} {\bibfnamefont {D.~N.~C.}\ \bibnamefont {{Lin}}},\ }\href {\doibase 10.3847/1538-3881/ab3230} {\bibfield  {journal} {\bibinfo  {journal} {\aj}\ }\textbf {\bibinfo {volume} {158}},\ \bibinfo {eid} {124} (\bibinfo {year} {2019})},\ \Eprint {http://arxiv.org/abs/1810.12925} {arXiv:1810.12925 [astro-ph.GA]} \BibitemShut {NoStop}%
\bibitem [{\citenamefont {{Goicoechea}}\ \emph {et~al.}(2019)\citenamefont {{Goicoechea}}, \citenamefont {{Santa-Maria}}, \citenamefont {{Bron}}, \citenamefont {{Teyssier}}, \citenamefont {{Marcelino}}, \citenamefont {{Cernicharo}},\ and\ \citenamefont {{Cuadrado}}}]{2019A&A...622A..91G}%
  \BibitemOpen
  \bibfield  {author} {\bibinfo {author} {\bibfnamefont {J.~R.}\ \bibnamefont {{Goicoechea}}}, \bibinfo {author} {\bibfnamefont {M.~G.}\ \bibnamefont {{Santa-Maria}}}, \bibinfo {author} {\bibfnamefont {E.}~\bibnamefont {{Bron}}}, \bibinfo {author} {\bibfnamefont {D.}~\bibnamefont {{Teyssier}}}, \bibinfo {author} {\bibfnamefont {N.}~\bibnamefont {{Marcelino}}}, \bibinfo {author} {\bibfnamefont {J.}~\bibnamefont {{Cernicharo}}}, \ and\ \bibinfo {author} {\bibfnamefont {S.}~\bibnamefont {{Cuadrado}}},\ }\href {\doibase 10.1051/0004-6361/201834409} {\bibfield  {journal} {\bibinfo  {journal} {\aap}\ }\textbf {\bibinfo {volume} {622}},\ \bibinfo {eid} {A91} (\bibinfo {year} {2019})},\ \Eprint {http://arxiv.org/abs/1812.00821} {arXiv:1812.00821 [astro-ph.GA]} \BibitemShut {NoStop}%
\bibitem [{\citenamefont {{Cao}}\ \emph {et~al.}(2022)\citenamefont {{Cao}}, \citenamefont {{Qiu}}, \citenamefont {{Zhang}},\ and\ \citenamefont {{Li}}}]{2022ApJ...927..106C}%
  \BibitemOpen
  \bibfield  {author} {\bibinfo {author} {\bibfnamefont {Y.}~\bibnamefont {{Cao}}}, \bibinfo {author} {\bibfnamefont {K.}~\bibnamefont {{Qiu}}}, \bibinfo {author} {\bibfnamefont {Q.}~\bibnamefont {{Zhang}}}, \ and\ \bibinfo {author} {\bibfnamefont {G.-X.}\ \bibnamefont {{Li}}},\ }\href {\doibase 10.3847/1538-4357/ac4696} {\bibfield  {journal} {\bibinfo  {journal} {\apj}\ }\textbf {\bibinfo {volume} {927}},\ \bibinfo {eid} {106} (\bibinfo {year} {2022})},\ \Eprint {http://arxiv.org/abs/2112.14080} {arXiv:2112.14080 [astro-ph.GA]} \BibitemShut {NoStop}%
\bibitem [{\citenamefont {{Sofue}}(2022)}]{2022MNRAS.516..907S}%
  \BibitemOpen
  \bibfield  {author} {\bibinfo {author} {\bibfnamefont {Y.}~\bibnamefont {{Sofue}}},\ }\href {\doibase 10.1093/mnras/stac2243} {\bibfield  {journal} {\bibinfo  {journal} {\mnras}\ }\textbf {\bibinfo {volume} {516}},\ \bibinfo {pages} {907} (\bibinfo {year} {2022})},\ \Eprint {http://arxiv.org/abs/2208.02451} {arXiv:2208.02451 [astro-ph.GA]} \BibitemShut {NoStop}%
\bibitem [{\citenamefont {{Ginsburg}}\ \emph {et~al.}(2016)\citenamefont {{Ginsburg}}, \citenamefont {{Henkel}}, \citenamefont {{Ao}}, \citenamefont {{Riquelme}}, \citenamefont {{Kauffmann}}, \citenamefont {{Pillai}}, \citenamefont {{Mills}}, \citenamefont {{Requena-Torres}}, \citenamefont {{Immer}}, \citenamefont {{Testi}}, \citenamefont {{Ott}}, \citenamefont {{Bally}}, \citenamefont {{Battersby}}, \citenamefont {{Darling}}, \citenamefont {{Aalto}}, \citenamefont {{Stanke}}, \citenamefont {{Kendrew}}, \citenamefont {{Kruijssen}}, \citenamefont {{Longmore}}, \citenamefont {{Dale}}, \citenamefont {{Guesten}},\ and\ \citenamefont {{Menten}}}]{2016A&A...586A..50G}%
  \BibitemOpen
  \bibfield  {author} {\bibinfo {author} {\bibfnamefont {A.}~\bibnamefont {{Ginsburg}}}, \bibinfo {author} {\bibfnamefont {C.}~\bibnamefont {{Henkel}}}, \bibinfo {author} {\bibfnamefont {Y.}~\bibnamefont {{Ao}}}, \bibinfo {author} {\bibfnamefont {D.}~\bibnamefont {{Riquelme}}}, \bibinfo {author} {\bibfnamefont {J.}~\bibnamefont {{Kauffmann}}}, \bibinfo {author} {\bibfnamefont {T.}~\bibnamefont {{Pillai}}}, \bibinfo {author} {\bibfnamefont {E.~A.~C.}\ \bibnamefont {{Mills}}}, \bibinfo {author} {\bibfnamefont {M.~A.}\ \bibnamefont {{Requena-Torres}}}, \bibinfo {author} {\bibfnamefont {K.}~\bibnamefont {{Immer}}}, \bibinfo {author} {\bibfnamefont {L.}~\bibnamefont {{Testi}}}, \bibinfo {author} {\bibfnamefont {J.}~\bibnamefont {{Ott}}}, \bibinfo {author} {\bibfnamefont {J.}~\bibnamefont {{Bally}}}, \bibinfo {author} {\bibfnamefont {C.}~\bibnamefont {{Battersby}}}, \bibinfo {author} {\bibfnamefont {J.}~\bibnamefont {{Darling}}}, \bibinfo {author} {\bibfnamefont {S.}~\bibnamefont {{Aalto}}}, \bibinfo {author} {\bibfnamefont {T.}~\bibnamefont {{Stanke}}}, \bibinfo {author} {\bibfnamefont {S.}~\bibnamefont {{Kendrew}}}, \bibinfo {author} {\bibfnamefont {J.~M.~D.}\ \bibnamefont {{Kruijssen}}}, \bibinfo {author} {\bibfnamefont {S.}~\bibnamefont {{Longmore}}}, \bibinfo {author} {\bibfnamefont {J.}~\bibnamefont {{Dale}}}, \bibinfo {author} {\bibfnamefont {R.}~\bibnamefont {{Guesten}}}, \ and\ \bibinfo {author} {\bibfnamefont {K.~M.}\ \bibnamefont {{Menten}}},\ }\href {\doibase 10.1051/0004-6361/201526100} {\bibfield  {journal} {\bibinfo  {journal} {\aap}\ }\textbf {\bibinfo {volume} {586}},\ \bibinfo {eid} {A50} (\bibinfo {year} {2016})},\ \Eprint {http://arxiv.org/abs/1509.01583} {arXiv:1509.01583 [astro-ph.GA]} \BibitemShut {NoStop}%
\bibitem [{\citenamefont {{Henshaw}}\ \emph {et~al.}(2016)\citenamefont {{Henshaw}}, \citenamefont {{Longmore}}, \citenamefont {{Kruijssen}}, \citenamefont {{Davies}}, \citenamefont {{Bally}}, \citenamefont {{Barnes}}, \citenamefont {{Battersby}}, \citenamefont {{Burton}}, \citenamefont {{Cunningham}}, \citenamefont {{Dale}}, \citenamefont {{Ginsburg}}, \citenamefont {{Immer}}, \citenamefont {{Jones}}, \citenamefont {{Kendrew}}, \citenamefont {{Mills}}, \citenamefont {{Molinari}}, \citenamefont {{Moore}}, \citenamefont {{Ott}}, \citenamefont {{Pillai}}, \citenamefont {{Rathborne}}, \citenamefont {{Schilke}}, \citenamefont {{Schmiedeke}}, \citenamefont {{Testi}}, \citenamefont {{Walker}}, \citenamefont {{Walsh}},\ and\ \citenamefont {{Zhang}}}]{2016MNRAS.457.2675H}%
  \BibitemOpen
  \bibfield  {author} {\bibinfo {author} {\bibfnamefont {J.~D.}\ \bibnamefont {{Henshaw}}}, \bibinfo {author} {\bibfnamefont {S.~N.}\ \bibnamefont {{Longmore}}}, \bibinfo {author} {\bibfnamefont {J.~M.~D.}\ \bibnamefont {{Kruijssen}}}, \bibinfo {author} {\bibfnamefont {B.}~\bibnamefont {{Davies}}}, \bibinfo {author} {\bibfnamefont {J.}~\bibnamefont {{Bally}}}, \bibinfo {author} {\bibfnamefont {A.}~\bibnamefont {{Barnes}}}, \bibinfo {author} {\bibfnamefont {C.}~\bibnamefont {{Battersby}}}, \bibinfo {author} {\bibfnamefont {M.}~\bibnamefont {{Burton}}}, \bibinfo {author} {\bibfnamefont {M.~R.}\ \bibnamefont {{Cunningham}}}, \bibinfo {author} {\bibfnamefont {J.~E.}\ \bibnamefont {{Dale}}}, \bibinfo {author} {\bibfnamefont {A.}~\bibnamefont {{Ginsburg}}}, \bibinfo {author} {\bibfnamefont {K.}~\bibnamefont {{Immer}}}, \bibinfo {author} {\bibfnamefont {P.~A.}\ \bibnamefont {{Jones}}}, \bibinfo {author} {\bibfnamefont {S.}~\bibnamefont {{Kendrew}}}, \bibinfo {author} {\bibfnamefont {E.~A.~C.}\ \bibnamefont {{Mills}}}, \bibinfo {author} {\bibfnamefont {S.}~\bibnamefont {{Molinari}}}, \bibinfo {author} {\bibfnamefont {T.~J.~T.}\ \bibnamefont {{Moore}}}, \bibinfo {author} {\bibfnamefont {J.}~\bibnamefont {{Ott}}}, \bibinfo {author} {\bibfnamefont {T.}~\bibnamefont {{Pillai}}}, \bibinfo {author} {\bibfnamefont {J.}~\bibnamefont {{Rathborne}}}, \bibinfo {author} {\bibfnamefont {P.}~\bibnamefont {{Schilke}}}, \bibinfo {author} {\bibfnamefont {A.}~\bibnamefont {{Schmiedeke}}}, \bibinfo {author} {\bibfnamefont {L.}~\bibnamefont {{Testi}}}, \bibinfo {author} {\bibfnamefont {D.}~\bibnamefont {{Walker}}}, \bibinfo {author} {\bibfnamefont {A.}~\bibnamefont {{Walsh}}}, \ and\ \bibinfo {author} {\bibfnamefont {Q.}~\bibnamefont {{Zhang}}},\ }\href {\doibase 10.1093/mnras/stw121} {\bibfield  {journal} {\bibinfo  {journal} {\mnras}\ }\textbf {\bibinfo {volume} {457}},\ \bibinfo {pages} {2675} (\bibinfo {year} {2016})},\ \Eprint {http://arxiv.org/abs/1601.03732} {arXiv:1601.03732 [astro-ph.GA]} \BibitemShut {NoStop}%
\bibitem [{\citenamefont {{Christopher}}\ \emph {et~al.}(2005)\citenamefont {{Christopher}}, \citenamefont {{Scoville}}, \citenamefont {{Stolovy}},\ and\ \citenamefont {{Yun}}}]{2005ApJ...622..346C}%
  \BibitemOpen
  \bibfield  {author} {\bibinfo {author} {\bibfnamefont {M.~H.}\ \bibnamefont {{Christopher}}}, \bibinfo {author} {\bibfnamefont {N.~Z.}\ \bibnamefont {{Scoville}}}, \bibinfo {author} {\bibfnamefont {S.~R.}\ \bibnamefont {{Stolovy}}}, \ and\ \bibinfo {author} {\bibfnamefont {M.~S.}\ \bibnamefont {{Yun}}},\ }\href {\doibase 10.1086/427829} {\bibfield  {journal} {\bibinfo  {journal} {\apj}\ }\textbf {\bibinfo {volume} {622}},\ \bibinfo {pages} {346} (\bibinfo {year} {2005})}\BibitemShut {NoStop}%
\bibitem [{\citenamefont {{Requena-Torres}}\ \emph {et~al.}(2012)\citenamefont {{Requena-Torres}}, \citenamefont {{G{\"u}sten}}, \citenamefont {{Wei{\ss}}}, \citenamefont {{Harris}}, \citenamefont {{Mart{\'i}n-Pintado}}, \citenamefont {{Requena}}, \citenamefont {{Stutzki}},\ and\ \citenamefont {{Zinnecker}}}]{2012A&A...542L..21R}%
  \BibitemOpen
  \bibfield  {author} {\bibinfo {author} {\bibfnamefont {M.~A.}\ \bibnamefont {{Requena-Torres}}}, \bibinfo {author} {\bibfnamefont {R.}~\bibnamefont {{G{\"u}sten}}}, \bibinfo {author} {\bibfnamefont {A.}~\bibnamefont {{Wei{\ss}}}}, \bibinfo {author} {\bibfnamefont {A.~I.}\ \bibnamefont {{Harris}}}, \bibinfo {author} {\bibfnamefont {J.}~\bibnamefont {{Mart{\'i}n-Pintado}}}, \bibinfo {author} {\bibfnamefont {S.}~\bibnamefont {{Requena}}}, \bibinfo {author} {\bibfnamefont {J.}~\bibnamefont {{Stutzki}}}, \ and\ \bibinfo {author} {\bibfnamefont {H.}~\bibnamefont {{Zinnecker}}},\ }\href {\doibase 10.1051/0004-6361/201219068} {\bibfield  {journal} {\bibinfo  {journal} {\aap}\ }\textbf {\bibinfo {volume} {542}},\ \bibinfo {eid} {L21} (\bibinfo {year} {2012})}\BibitemShut {NoStop}%
\end{thebibliography}%
\end{document}